\documentclass[journal=jpclcd,manuscript=article]{achemso}
\setkeys{acs}{articletitle=true}

\usepackage[version=3]{mhchem} 

\usepackage{amsmath}
\usepackage{amsfonts}
\usepackage{amssymb}
\usepackage{graphicx}
\usepackage{dcolumn}
\usepackage{bm}
\usepackage{subcaption}
\usepackage{algorithm} 
\usepackage{algpseudocode} 
\usepackage{xr}
\usepackage{hyperref}
\usepackage{braket}
\usepackage{xcolor}

\makeatletter
\newcommand*{\addFileDependency}[1]{
  \typeout{(#1)}
  \@addtofilelist{#1}
  \IfFileExists{#1}{}{\typeout{No file #1.}}
}
\makeatother

\usepackage[capitalize]{cleveref}
\crefname{figure}{Fig.}{Figs.}
\Crefname{figure}{Figure}{Figures}
\crefname{table}{Tab.}{Tabs.}
\Crefname{table}{Table}{Tables}
\crefname{equation}{Eq.}{Eqs.}
\Crefname{equation}{Equation}{Equations}
\crefname{section}{Sec.}{Secs.}
\Crefname{section}{Section}{Sections}
\usepackage{xcolor}

\usepackage{array}
\newcommand{\PreserveBackslash}[1]{\let\temp=\\#1\let\\=\temp}
\newcolumntype{C}[1]{>{\PreserveBackslash\centering}p{#1}}
\newcolumntype{R}[1]{>{\PreserveBackslash\raggedleft}p{#1}}
\newcolumntype{L}[1]{>{\PreserveBackslash\raggedright}p{#1}}

\usepackage{multirow}

\author{Fabijan Pavo\v{s}evi\'{c}}
\email{fpavosevic@gmail.com, fabijan.pavosevic@algorithmiq.fi}
\affiliation{Algorithmiq Ltd., Kanavakatu 3C, FI-00160 Helsinki, Finland}
\alsoaffiliation{Center for Computational Quantum Physics, Flatiron Institute, 162 5th Ave., New York, 10010  NY,  USA}

\author{Ivano Tavernelli}
\affiliation{IBM Quantum, IBM Research – Zürich, 8803 Rüschlikon, Switzerland}

\author{Angel Rubio}
\alsoaffiliation{Center for Computational Quantum Physics, Flatiron Institute, 162 5th Ave., New York, 10010  NY,  USA}
\affiliation{Max Planck Institute for the Structure and Dynamics of Matter and
Center for Free-Electron Laser Science \& Department of Physics,
Luruper Chaussee 149, 22761 Hamburg, Germany}

\title[]
  {Spin-Flip Unitary Coupled Cluster Method: Toward Accurate Description of Strong Electron Correlation on Quantum Computers}

\begin{document}



\begin{tocentry}
\begin{figure}[H]
	\begin{center}
		\includegraphics[width=1.7in]{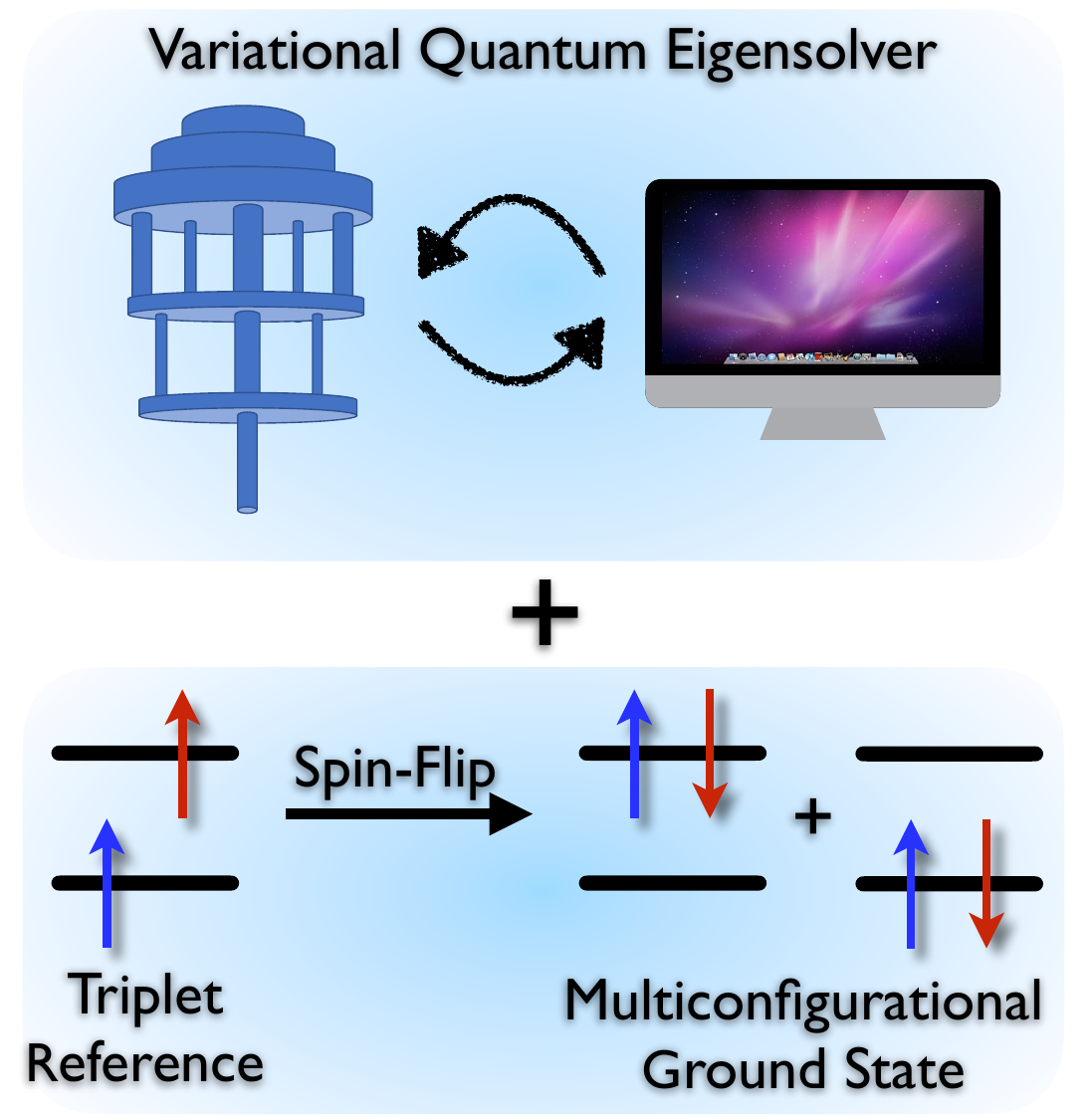}

	\end{center}
\end{figure}
\end{tocentry}

\begin{abstract}
Quantum computers have emerged as a promising platform to simulate the strong electron correlation that is crucial to catalysis and photochemistry. However, owing to the choice of a trial wave function employed in the popular hybrid quantum-classical variational quantum eigensolver (VQE) algorithm, the accurate simulation is restricted to certain classes of correlated phenomena. Herein, we combine the spin-flip (SF) formalism with the unitary coupled cluster with singles and doubles (UCCSD) method via the quantum equation-of-motion (qEOM) approach to allow for an efficient simulation of a large family of strongly correlated problems. In particular, we show that the developed qEOM-SF-UCCSD/VQE method outperforms its UCCSD/VQE counterpart for simulation of the $cis$-$trans$ isomerization of ethylene and the automerization of cyclobutadiene. The predicted qEOM-SF-UCCSD/VQE barrier heights for these two problems are in a good agreement with the experimentally determined values. The methodological developments presented herein will further stimulate investigation of this approach for the simulation of other types of correlated/entangled phenomena on a quantum computer.
\end{abstract}

\maketitle


Simulation of strongly correlated systems is at the heart of chemical and materials science. Development of techniques that enable accurate and efficient description of the strong electron correlation in molecules, solids, and biocomplexes is a key for controlling and designing systems with tailored functionalities. Owing to their very complex electronic structure that arises due to multiple close in energy electronic states~\cite{lyakh2012multireference}, many popular and computationally efficient single-reference quantum chemistry models, such as Kohn-Sham density functional theory (KS-DFT) methods~\cite{Sham65_1133} and coupled cluster methods (CC)~\cite{bartlett2007coupled}, delivers qualitatively incorrect results for such systems. The accurate description of the strong electron correlation requires the multireference quantum chemistry methods~\cite{Szalay2011}. In these methods, the electronic wave function is expanded in terms of weighted electronic configurations, but because of the factorial scaling of the configurational space, their applicability is limited to small molecules with up to a couple dozen of electrons. 

Quantum algorithms~\cite{bauer2020quantum} on the other hand, offer a path for dramatically reducing this unfavorable factorial scaling when the wave function is prepared on quantum devices~\cite{cao2019quantum}. One such algorithm is the quantum phase estimation (QPE)~\cite{aspuru2005simulated}, which under certain assumptions exhibits an exponential speedup for evaluation of eigenvalues of the molecular Hamiltonian compared to the exact classical algorithm. However, its implementation requires circuit depths~\cite{bauer2020quantum} that are beyond abilities of the currently available noisy near-term quantum devices~\cite{preskill2018quantum}. In contrast, the hybrid quantum-classical variational quantum eigensolver (VQE) algorithm~\cite{peruzzo2014variational} operates with relatively low circuit depths making it suitable for the present-day quantum devices. Due to this prominent and favourable feature, the VQE algorithm has been successfully implemented on different quantum computing architectures for simulation of small molecules~\cite{peruzzo2014variational,o2016scalable,shen2017quantum,hempel2018quantum}. Within the VQE algorithm, preparation of a parametrized trial wave function and the ground-state energy estimation are performed on a quantum device, whereas the wave function parameters are optimized on a classical computer via the variational principle~\cite{peruzzo2014variational}. 

In practice, the choice of the parametrized trial wave function determines the accuracy, performance, and efficiency of the VQE algorithm. The original implementation of the VQE algorithm was making use of the chemically inspired unitary coupled cluster with singles and doubles (UCCSD) method~\cite{peruzzo2014variational}. Despite the fact that the UCCSD method suffers from the relatively deep circuits, it exhibits several important features: unitarity, variational character, possibility of systematic improve of the Ansatz, and became therefore a main source of inspiration for development of the other more efficient ansatzes~\cite{ryabinkin2018qubit,lee2018generalized,grimsley2019adaptive,sokolov2020quantum,fitzpatrick2022self}. Yet another advantage of the unitary coupled cluster method is a good performance for simulation of selected strongly correlated systems. In particular, most demonstrations indicate a good performance in simulating the covalent bond-breaking problems~\cite{cooper2010benchmark,anand2022quantum}, which is a classic example of the strong electron correlation. However, due to the single-reference nature of the wave function, the UCCSD method exhibits a poor performance in describing the $\text{Be} + \text{H}_2 \rightarrow \text{BeH}_2$ reaction around the transition structure where the strong electron correlation is prominent~\cite{evangelista2011alternative}. 
Inclusion of the triple excitations systematically improves the accuracy of the method, albeit at an increasing cost due to additional variational parameters~\cite{evangelista2011alternative}. This has prompted development of different variants of the multireference unitary coupled cluster (MR-UCC) methods that employ the complete active space self-consistent field (CASSCF) reference instead of the Hartree-Fock (HF) reference~\cite{greene2021generalized}. The developed MR-UCC methods display a significant improvement over their single-reference counterparts for modeling of the aforementioned reaction~\cite{sugisaki2022variational}. 

The obvious drawback of the MR-UCC methods is that preparation of the CASSCF reference wave function scales exponentially with the active space size, making the use of the VQE algorithm for simulation of large strongly correlated molecular systems on quantum devices unpractical. Therefore, it is of the utmost importance to develop the UCC method and its variants suitable for simulations of such strongly correlated systems that does not suffer from the CASSCF bottleneck. Moreover, it is also important to investigate the performance of the UCCSD method and its improved variants for other types of chemical reactions and processes where strong electron correlation plays a crucial role. These other examples include the rotation around double chemical bonds, the process responsible for vision as well as for photo-switching~\cite{kiser2014chemistry,tochitsky2018restoring}. Another example is the automerization of the cyclobutadiene molecule where the transition state exhibits a significant strongly correlated character~\cite{lyakh2012multireference}.

A way to avoid the multireference approach can be achieved with the spin-flip (SF) formalism. The significant advantage of the SF formalism is that it allows for accurate treatment of the strongly correlated systems within the single-reference framework~\cite{krylov2001size,casanova2020spin}. Within the SF formalism, the target multiconfigurational wave function is obtained by the action of the spin-flip excitation operator on a high-spin reference state via the linear response~\cite{shao2003spin} or equation-of-motion (EOM) formalisms~\cite{krylov2001size,shao2003spin}. Over the past two decades the SF formalism has been combined with both DFT~\cite{shao2003spin} and wave function methods~\cite{krylov2001size} and was successfully applied to study hard-problems in chemistry, such as, bond-breaking processes, radical systems, conical intersections, and excited-state processes~\cite{casanova2020spin}. Particularly effective is the EOM-SF-CCSD method~\cite{krylov2001size}, however due to a nonvariational nature of the CCSD method, certain problems can arise~\cite{thomas2021complex}. On the other hand, the UCCSD method is variational, but due to non-truncation of the Baker-Campbell-Hausdorff expansion of the similarity transformed UCCSD Hamiltonian~\cite{taube2006new}, this method has an exponential cost on a classical computer. On a quantum computer, the solution of the UCCSD method in conjunction with the VQE algorithm can be achieved at the polynomial cost under well-defined conditions~\cite{peruzzo2014variational}. 

To facilitate an accurate simulation of different classes of strongly correlated problems on the noisy near-term quantum devices and to avoid the CASSCF bottleneck as present in the MR-UCC methods, herein we combine the SF formalism with the variational UCCSD/VQE method via the quantum equation-of-motion (qEOM) method~\cite{ollitrault2020quantum}. The performance and accuracy of the developed qEOM-SF-UCCSD/VQE method is verified and benchmarked on the prototype $cis$-$trans$ isomerization reaction in ethylene and on the automerization reaction of cyclobutadiene. The developments and analysis demonstrated herein highlight the efficiency and accuracy of the qEOM-SF-UCCSD/VQE method for treatment of different classes of strongly correlated problems in molecular chemistry on the present-day quantum computers. Moreover, this work lays the foundation for development of other quantum-hardware efficient methods within the spin-flip formalism that can be extended to larger molecular systems and solids via the embedding techniques~\cite{chulhai2018projection,rossmannek2023quantum}.


In the UCC approach, the wave function ansatz is given by 
\begin{equation}
    \label{eqn:UCC}
    |\Psi_{\text{UCC}}\rangle=e^{\hat{T}-\hat{T}^\dagger}|0\rangle
\end{equation}
where $|0\rangle$ is the uncorrelated reference HF wave function, and $\hat{T}=\theta_{\mu}a^{\mu}$ is the excitation cluster operator that accounts for the correlation effects between electrons. The cluster operator is expressed in terms of set of single, double, and higher excitation operators, $a^\mu=a_\mu^{\dagger}=\{a_{i}^{a},a_{ij}^{ab},...\}$, where the general electronic excitation operator $a_{p_1p_2...p_n}^{q_1q_2...q_n}=a_{q_1}^{\dagger}a_{q_2}^{\dagger}...a_{q_n}^{\dagger}a_{p_n}...a_{p_2}a_{p_1}$ is defined as a string of fermionic creation ($a^{\dagger}$) and annihilation ($a$) operators, and $\mu$ represents the excitation rank. Furthermore, $p,q,r,s,...$ indices denote general electronic spin orbitals, $i,j,k,l,...$ denote occupied electronic spin orbitals, $a,b,c,d,...$ denote unoccupied electronic spin orbitals. The  amplitudes $\theta_{\mu}$ are unknown wave function parameters that are determined within the VQE algorithm by variationally minimizing the energy functional
\begin{equation}
    \label{eqn:VQE-Energy-Funtional}
     E_{\text{UCC/VQE}}=\underset{\theta}{\text{min}} \langle\Psi_{\text{UCC}}|\hat{H}|\Psi_{\text{UCC}}\rangle
\end{equation}
which gives the ground state energy $E_{\text{UCC/VQE}}$~\cite{peruzzo2014variational}. Here, $\hat{H} = h^p_q a^q_p + \frac{1}{2}g^{pq}_{rs}a^{rs}_{pq}$ is the electronic Hamiltonian within the Born-Oppenheimer approximation (going beyond this approximation is also possible~\cite{pavošević2020multicomponent,pavošević2018multicomponent,pavošević2021multicomponent,pavosevic2020frequency,pavosevic2021multicomponent}) expressed using the second-quantized formalism. Additionally, $h^p_q=\langle q |\hat{h}|p \rangle$ and $g^{pq}_{rs}=\langle rs|pq \rangle$ are matrix elements of the one-electron core Hamiltonian and tensor elements of the two-electron Coulomb repulsion integrals, respectively, in the spin orbital basis. 

The excitation energies at the UCC level can be obtained within the quantum equation-of-motion (qEOM-UCC) formalism~\cite{ollitrault2020quantum}. The excitation energy $\Delta E_I$ of the $I$-th excited state is obtained from 
\begin{equation}
    \label{eqn:EOM-Excitation}
    \Delta E_I=\frac{\langle\Psi_{\text{UCC}}|[\hat{O}_I,\hat{H},\hat{O}_I^{\dagger}]|\Psi_{\text{UCC}}\rangle}{\langle\Psi_{\text{UCC}}|[\hat{O}_I,\hat{O}_I^{\dagger}]|\Psi_{\text{UCC}}\rangle}
\end{equation}
where commutators are defined as $[\hat{O}_I,\hat{O}_I^{\dagger}]=\hat{O}_I\hat{O}_I^{\dagger}-\hat{O}_I^{\dagger}\hat{O}_I$ and $[\hat{O}_I,\hat{H},\hat{O}_I^{\dagger}]=([[\hat{O}_I,\hat{H}],\hat{O}_I^{\dagger}]+[\hat{O}_I,[\hat{H},\hat{O}_I^{\dagger}]])/2$, and
$\hat{O}_I^{\dagger}=\sum_{\mu}\big[X_{\mu}(I)a^{\mu}-Y^{\mu}(I)a_{\mu}\big]$ is the electron excitation operator. The excitation energies and the excited state wave function parameters $X_I$ and $Y_I$ are determined by solving the following generalized eigenvalue problem 
\begin{equation}
    \label{eqn:EOM-Equation}
    \begin{pmatrix}
        \mathbf{A} & \mathbf{B}\\
        \mathbf{B}^* & \mathbf{A}^*
    \end{pmatrix}
    \begin{pmatrix}
        \mathbf{X}_I\\
        \mathbf{Y}_I
    \end{pmatrix}
    =\Delta E_I
    \begin{pmatrix}
        \mathbf{C} & \mathbf{D}\\
        -\mathbf{D}^* & -\mathbf{C}^*
    \end{pmatrix}
    \begin{pmatrix}
        \mathbf{X}_I\\
        \mathbf{Y}_I
    \end{pmatrix}
\end{equation}
with the matrix elements defined as
\begin{equation}
    \label{eqn:QED-EOM-Mat-Elem}
        \begin{aligned}
            A_{\mu,\nu}&=\langle\Psi_{\text{UCC}}|[a_{\mu},\hat{H},a^{\nu}]|\Psi_{\text{UCC}}\rangle \newline\\
            B_{\mu,\nu}&=-\langle\Psi_{\text{UCC}}|[a_{\mu},\hat{H},a_{\nu}]|\Psi_{\text{UCC}}\rangle \newline\\
            C_{\mu,\nu}&=\langle\Psi_{\text{UCC}}|[a_{\mu},a^{\nu}]|\Psi_{\text{UCC}}\rangle \newline\\
            D_{\mu,\nu}&=-\langle\Psi_{\text{UCC}}|[a_{\mu},a_{\nu}]|\Psi_{\text{UCC}}\rangle 
        \end{aligned}
\end{equation}
The nature of the $a^{\mu}/a_{\mu}$ operators that take part to the excitation operator $\hat{O}^{\dagger}$ defines different qEOM models~\cite{asthana2023quantum}. Therefore, in case where $a^{\mu}=a_{\mu}^{\dagger}=\{a_{i}^{a},a_{ij}^{ab},...\}$ is defined in terms of particle-conserving operators produces electronic excitation energies, whereas in cases where these operators are given in terms of particle-non-conserving operators $a^{\mu}=a_{\mu}^{\dagger}=\{a_{i},a_{ij}^{a},...\}$ and $a^{\mu}=a_{\mu}^{\dagger}=\{a^{a},a_{i}^{ab},...\}$, allows calculations of ionization potentials and electron affinities, respectively~\cite{asthana2023quantum}. In general, the operators $a^{\mu}/a_{\mu}$ in $\hat{O}^{\dagger}$ are defined in terms of spin-conserving operators~\cite{ollitrault2020quantum,pavosevic2021multicomponent,pavosevic2021polaritonic,asthana2023quantum,kim2023two}. In case of the spin-flip (SF) formalism, these operators are defined in terms of excitation operators that flips the spin of one electron ($\alpha\rightarrow\beta$)~\cite{krylov2001size,casanova2020spin} from the reference UCCSD
wave function. As a result, when such spin-flip excitation operator acts on the high-spin reference state (in this work we consider triplet state), it produces the multiconfigurational low-spin target state (singlet state). The resulting spin-flip approach is coined as the qEOM-SF-UCC method.

For practical purposes, the electronic excitation operators $a^{\mu}=a_{\mu}^{\dagger}=\{a_{i}^{a},a_{ij}^{ab},...\}$ that appear in the cluster operator $\hat{T}$ and in the qEOM operator $\hat{O}^{\dagger}$, are truncated at singles and doubles level. Such truncation defines the unitary coupled cluster with singles and doubles (UCCSD) method and the quantum equation-of-motion unitary coupled cluster with singles and doubles (qEOM-UCCSD) excited-state method as well as its spin-flip counterpart (qEOM-SF-UCCSD) suitable for quantum devices. For simplicity, we will refer to this spin-flip method as SF-UCCSD.


The UCCSD, qEOM-UCCSD, and SF-UCCSD methods were implemented in an in-house version of the \textsf{Psi4NumPy} quantum chemistry software~\cite{smith2018psi4numpy}. These methods were formulated in the qubit basis making them suitable for implementations on quantum devices within the VQE algorithm. This implementation follows the usual steps of the VQE procedure~\cite{peruzzo2014variational}; in the first step, the second-quantized Hamiltonian is formulated in the spin orbital basis using the HF spin orbitals calculated with \textsf{Psi4NumPy}. In the second step, the second-quantized Hamiltonian and the second-quantized excitation operators are converted from the fermionic into the qubit basis via the Jordan-Wigner mapping~\cite{jordan1993algebraic} ($a^{\dagger}_p=1/2(X^p-iY^p)\otimes_{q<p}Z^q$, where $X,Y,Z$ are the Pauli matrices) as implemented in \textsf{OpenFermion}~\cite{mcclean2020openfermion}. Following this step, the energy in Eq.~\eqref{eqn:VQE-Energy-Funtional} is variationally optimized by employing the Broyden-Fletcher-Goldfarb-Shanno algorithm as implemented in \textsf{SciPy}~\cite{virtanen2020scipy}. This step provides the optimal UCCSD wave function parameters $\theta$ that minimizes the ground state energy. In the last step, the optimal UCCSD wave function is used to construct Eq.~\eqref{eqn:EOM-Equation} whose diagonalization provides the excitation energies. In the present study, we employ the quantum state vector simulation rather than relying on statistical sampling of the noisy quantum measurements. Because the focus of this work is benchmarking of the developed SF-UCCSD method, the state vector simulation is more suitable for our purposes. In the following, we discuss the comparison of the performance of the UCCSD and SF-UCCSD methods for simulation of two different systems in which the strong electron correlation is significant. 

\begin{figure*}[ht!]
  \centering
  \includegraphics[width=3.25in]{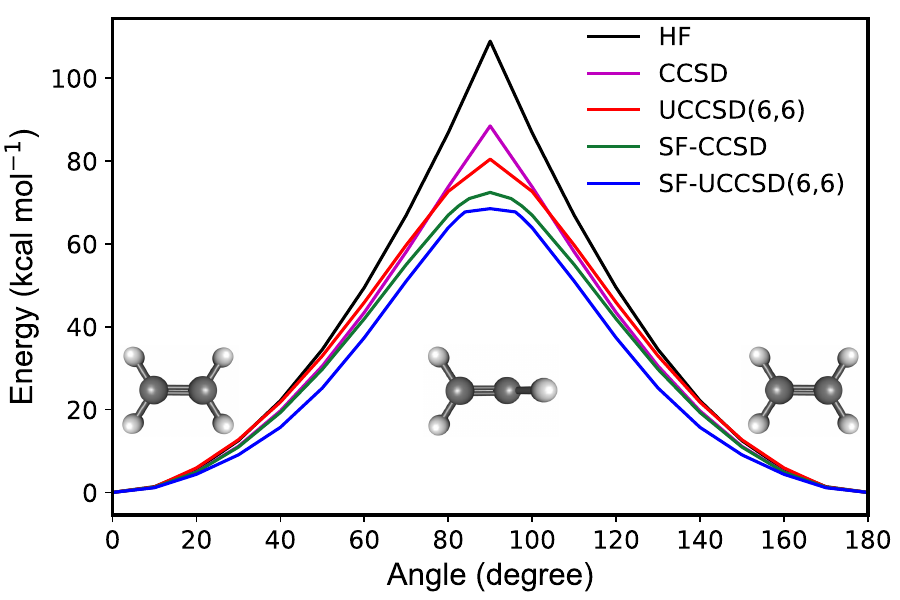}
  \caption{Potential energy curve as a function of the twisting angle in the ethylene molecule calculated with the HF (black), CCSD (magenta), UCCSD (red), SF-CCSD (green), and SF-UCCSD (blue) methods employing the cc-pVDZ basis set. The UCCSD and SF-UCCSD calculations are performed with active space comprised from 6 electrons in 6 orbitals. The plot also contains images of equilibrium (0$^\circ$ and 180$^\circ$) and twisted (90$^\circ$) geometries of the ethylene molecule.}
  \label{fig:figure1}
\end{figure*}

\begin{table}[]

\caption{Reaction energy barrier\textsuperscript{a} (in kcal/mol) for the torsion of the ethylene molecule calculated with different wave function methods employing the cc-pVDZ basis set. }

\begin{tabular}{C{3cm}|C{3.5cm}}
\hline
 & Barrier (kcal/mol) \\ \hline\hline
HF & 108.9 \\ \hline
CCSD &  88.46\\ \hline
CCSD(6,6) & 80.7 \\ \hline
CCSD(8,8) & 78.0 \\ \hline
CCSD(10,10) & 79.4 \\ \hline
UCCSD(6,6) & 80.4 \\ \hline
UCCSD(8,8) & 78.0 \\ \hline
UCCSD(10,10) & 79.4 \\ \hline
SF-CCSD & 72.4 \\ \hline
SF-CCSD(6,6) & 68.5 \\ \hline
SF-UCCSD(6,6) & 68.4 \\ \hline
Experiment~\cite{douglas1955kinetics,wallace1989torsional} & 60-65 \\ \hline\hline
\end{tabular}

\raggedright \textsuperscript{a}\small Calculated as the energy difference between the twisted (90$^\circ$) and the equilibrium (0$^\circ$) geometry.\\

\label{table:table1}
\end{table}

As our first example, we study the rotation around the double bond in the ethylene molecule which is a prototypical example of the $cis$-$trans$ isomerization process involved in vision and light-switching~\cite{kiser2014chemistry,tochitsky2018restoring}. Due to its complicated electronic structure, this process poses a challenge for the single-reference wave function methods. The single-reference methods are capable of accurately describing the electronic structure around the equilibrium geometry of the ethylene molecule where the wave function is dominated by a single $(\pi)^2$ configuration and where the energy gap between $\pi$ bonding orbital and $\pi^*$ anti-bonding orbital is significant. However, as the torsion around the double bond progresses (see Figure~\ref{fig:figure1}), the energy gap between the $\pi$ and $\pi^*$ orbitals decreases and importance of the $(\pi^*)^2$ configuration increases. Finally, at the barrier which corresponds to a region on the potential energy surface (PES) where the dihedral angle is 90$^{\circ}$, the $\pi$ and $\pi^*$ orbitals become energetically degenerate and the $(\pi)^2$ and $(\pi^*)^2$ configurations become equally important. Due to unbalanced treatment of the $(\pi)^2$ and $(\pi^*)^2$ configurations, the single-reference methods fail to describe the correct behaviour when the dihedral angle is 90$^{\circ}$.   

Figure~\ref{fig:figure1} shows the potential energy curve as a function of the twisting angle in the ethylene molecule calculated with different wave function methods. The reported calculations employ the cc-pVDZ basis set~\cite{dunning1989gaussian} and are performed on the geometry optimized at the CCSD/cc-pVDZ level of theory using the Q-Chem quantum chemistry software.~\cite{epifanovsky2021software} Along the PES scan, only the corresponding dihedral (twist) angle is changed, while all other degrees of freedom are kept frozen. As shown in  Fig.~\ref{fig:figure1}, the HF method (black curve) exhibits an unphysical cusp when the twisting dihedral angle is 90$^{\circ}$ because the HF wave function consists of only one configuration. The conventional single-reference CCSD method (magenta curve) even though it includes additional configurations, has the same spurious cusp for twisted geometry due to the fact that the other configurations are not treated on equal footing as the reference $(\pi)^2$ configuration. Numerical values of the reaction barrier for this process calculated with different methods are collected in Table~\ref{table:table1}. Even though the energy barrier calculated with the CCSD method is lower by 20~kcal/mol relative to the HF method, the CCSD method overestimates the experimentally determined reaction barrier by more than 20~kcal/mol~\cite{douglas1955kinetics,wallace1989torsional}.

Next, we discuss the performance of the UCCSD/VQE method. Because the state-vector implementation of the UCCSD method has an inherent exponential scaling with the system/basis set size, we have restricted our calculations to a relatively small active space consisting of 6 electrons in 6 orbitals closest to the Fermi level. Figure~\ref{fig:figure1} indicates that the UCCSD(6,6) (red curve) method has the same problem with the nonphysical cusp as the CCSD method. Moreover, the CCSD(6,6) method in the same active space delivers nearly identical results to the UCCSD(6,6) method (see Table~\ref{table:table1}), and for clarity the CCSD(6,6)  results are not included in Fig.~\ref{fig:figure1}. Therefore, we conclude that the UCCSD/cc-pVDZ with a large active space will maintain the same unphysical cusp and will exhibit a similar performance to the CCSD method for simulation of such problem. To confirm this conjecture, we performed both CCSD and UCCSD calculations using two larger active spaces made of 8 electrons in 8 orbitals, and 10 electrons in 10 orbitals respectively; the results remain nearly identical in both cases. We can therefore conclude that the UCCSD/cc-pVDZ method for this class of problem suffers from the same issue as the conventional single-reference CCSD/cc-pVDZ method. 

Next, we focus on the performance of the spin-flip methods in dealing with the cusp problems discussed above. Unlike the HF, CCSD, and UCCSD methods, the conventional electronic structure SF-CCSD (green curve) method produces a smooth curve as shown in Fig.~\ref{fig:figure1}. We note that for the purpose of this work we have also implemented the conventional electronic structure SF-CCSD method. The calculated SF-CCSD barrier is 72.4~kcal/mol, which is in a good agreement with the experimental values. Note that this calculation does not account for geometry relaxation of the transition state structure. Lastly, the SF-UCCSD(6,6) method also produces a smooth cuspless curve with the barrier height of 68.5~kcal/mol. The discrepancy between the SF-UCCSD(6,6) and SF-CCSD is due to the choice of the active space. This is confirmed by running the SF-CCSD(6,6) method using the same active space and the results are nearly identical as shown in Table~\ref{table:table1}. Although the energy profile is too flat at the top of the barrier, an increase of the active space will improve the  performance of the SF-UCCSD method and therefore the shape of the curve.

\begin{figure*}[ht!]
  \centering
  \includegraphics[width=3.25in]{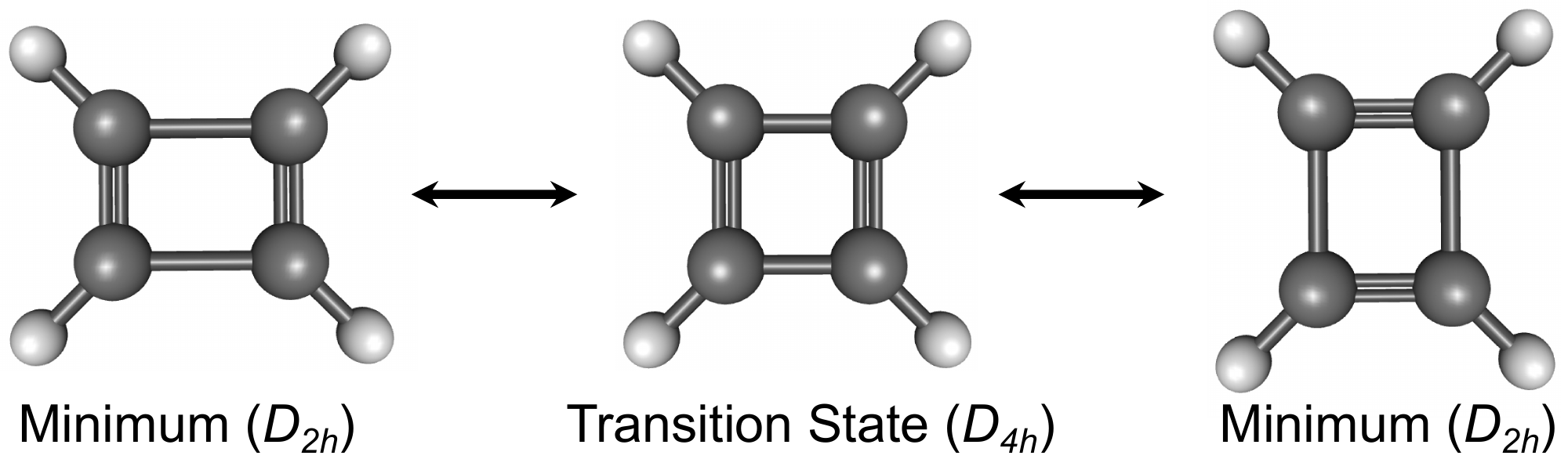}
  \caption{The automerization of the cyclobutadiene molecule.}
  \label{fig:figure2}
\end{figure*}

As our second example, we study the automerization of  cyclobutadiene (see Fig.~\ref{fig:figure2}). This process is yet another strongly correlated problem for which the single-reference methods struggle to describe the correct automerization barrier height. Therefore, this system is excellent for testing the validity of various methods when the strong electronic correlation is significant. The equilibrium rectangular ($D_{2h}$) geometry is well described by the single-reference methods, however, the transition state square ($D_{4h}$) geometry, where the two reference configurations becomes equally important, exhibits a significant strongly correlated character.

\begin{table}[!htb]

\caption{Barrier heights for the automerization of cyclobutadiene\textsuperscript{a} (in kcal/mol) calculated with different wave function methods employing the cc-pVDZ basis set.}

\begin{tabular}{C{3cm}|C{3.5cm}}
\hline
 & Barrier (kcal/mol) \\ \hline\hline
HF & 30.6 \\ \hline
CCSD &  20.9\\ \hline
CCSD(4,4) & 15.0 \\ \hline
CCSD(10,10) & 17.9 \\ \hline
UCCSD(4,4) & 15.0 \\ \hline
UCCSD(10,10) & 17.9 \\ \hline
SF-UCCSD(4,4) & 7.6 \\ \hline
Experiment~\cite{whitman1982limits} & 1.6-10 \\ \hline\hline
\end{tabular}

\raggedright \textsuperscript{a}\small Calculated as the energy difference between the transition state ($D_{4h}$) and the equilibrium ($D_{2h}$) geometry.\\

\label{table:table2}
\end{table}

Table~\ref{table:table2} summarizes the calculated automerization energy barriers of cyclobutadiene. All of the calculations were performed on geometries obtained from Ref.~\citenum{demel2006multireference} and employ the cc-pVDZ basis set~\cite{dunning1989gaussian}. The CCSD method greatly underestimates the importance of the doubly excited determinant and predicts the barrier height of 20.9~kcal/mol. Therefore, due to lack of inclusion of the static correlation, it greatly  overestimates the experimentally determined value that is in the range of 1.6-10~kcal/mol~\cite{whitman1982limits}. Similarly, the UCCSD(4,4) method with the active space comprised of 4 electrons in 4 $\pi$ orbitals, predicts the automerization barrier height of 15.0~kcal/mol. The identical value is determined with the CCSD(4,4) method using the same active space. Therefore, the two methods exhibit the equivalent behaviour for this type of system and fail to account for multiconfigurational wave function at the transition state geometry. To confirm this observation, we run the same calculations with an active space comprised of 10 electrons in 10 orbitals, which led to the same barrier as predicted by the CCSD(10,10) and UCCSD(10,10) calculations, as shown in Table~\ref{table:table2}. Lastly, the newly developed SF-UCCSD(4,4) approach predicts this barrier height at 7.6~kcal/mol which is in agreement with the experimental values. Therefore, the SF-UCCSD method is capable of accurately describing the region of the PES where the strong correlation is prominent. For comparison, we also note that various highly sophisticated multireference coupled cluster methods predict the barrier height to be 5-10~kcal/mol~\cite{lyakh2012multireference}.


In this work, we extend the applicability of the UCCSD/VQE approach for quantum computations to different strongly correlated problems by means of the spin-flip (SF) formalism. Herein, we show that the UCCSD/VQE method fails to accurately describe the $cis$-$trans$ isomerization process of ethylene and the automerization of cyclobutadiene. 
Due to its single-reference wave function character, the UCCSD method overestimates the corresponding barrier heights and exhibits nearly identical behaviour to the conventional electronic structure CCSD method for both cases ($cis$-$trans$ isomerization and automerization). 
On the other hand, the SF-UCCSD/VQE method presented herein, predicts the correct barrier heights, which are in a good agreement with experimentally determined values. 
The SF-UCCSD approach combines the UCCSD/VQE method and the modified qEOM algorithm in which the excitation operator allows for the spin flip of a single electron relative to the reference UCCSD wave function. 
In view of the quality of the results presented in this work with the proposed SF-UCCSD/VQE approach, together with its performance and simplicity, we argue that this approach may become the method of choice for extending the applicability of the UCCSD/VQE method to strongly correlated systems. 
Following the growing interest in the design of more efficient near-term quantum algorithms for quantum chemistry applications, the proposed spin-flip formalism can open up new frontiers in the search for improved quantum hardware-efficient algorithms in electronic structure calculations.  
More generally, we expect that our findings will shed new light into the application of the SF-UCCSD/VQE approach in the study of strongly correlated systems in general, and their role in catalysis and biological processes.

\begin{acknowledgement}
We acknowledge financial support from the Cluster of Excellence 'CUI: Advanced Imaging of Matter'- EXC 2056 - project ID 390715994 and SFB-925 "Light induced dynamics and control of correlated quantum systems" – project 170620586  of the Deutsche Forschungsgemeinschaft (DFG) and Grupos Consolidados (IT1453-22). We also acknowledge support from the Max Planck–New York Center for Non-Equilibrium Quantum Phenomena. The Flatiron Institute is a division of the Simons Foundation. This research was supported by the NCCR MARVEL, a National Centre of Competence in Research, funded by the Swiss National Science Foundation (grand number 205602). IBM, the IBM logo, and ibm.com are trademarks of International Business Machines Corp., registered in many jurisdictions worldwide. Other product and service names might be trademarks of IBM or other companies. The current list of IBM trademarks is available at \url{https://www.ibm.com/legal/copytrade}.
\end{acknowledgement}



\noindent\textbf{Conflict of interest}\\
The authors declare no conflict of interest.

\linespread{1}\selectfont
\bibliography{references}{}

\end{document}